\documentstyle[twoside,fleqn,espcrc2,epsf]{article}

\title{Hadron masses in QCD on a $16^3\times40$ lattice at $\beta=5.7$
       with $ma=0.01$}

\author{Dong Chen\address{Department of Physics, Columbia University,
	New York, NY 10027, USA}
        \thanks{This work was done in collaboration with Norman H. Christ,
		Robert D. Mawhinney, Igor Arsenin, Shailesh Chandrasekharan,
		Weonjong Lee and Decai Zhu and supported in part by the 
		Department of Energy.}}
       
\begin{document}

\def\thepage{CU--TP--664 \ \ \  hep-lat/9412069}
\thispagestyle{myheadings}

\begin{abstract}
We report on the hadron mass spectrum obtained on a $16^3 \times 40$ lattice
in full QCD at $\beta = 5.7$ using two flavors of staggered fermions 
with $m a = 0.01$.  We study the effective mass plateaus for different sized
sources.  Our mass results are slightly lighter than our earlier $16^3 \times
32$ calculation.  The Landau gauge $\Delta$ is quite different from 
the Coulomb gauge $\Delta$.
\end{abstract}

\maketitle

\section{INTRODUCTION}
The hadron mass spectrum is one of the fundamental quantities to calculate in
lattice QCD.  The Columbia group has done hadron mass calculations with
two flavors of staggered dynamical quarks for 
$m a = 0.01, 0.015, 0.02, 0.025$ on a $16^3 \times 32$ volume and 
$m a = 0.01, 0.025$ on $32^4$, both at $\beta = 5.7$ 
\cite{Hong91,Brown91,Wendy93}.  We also investigated 
$m a = 0.004$ on $16^3 \times 32$ at $\beta = 5.48$ \cite{Chen94}.

In this note, we report a new study on a $16^3 \times 40$ lattice with
$m a = 0.01$ at $\beta = 5.7$.  The increased time dimension from $N_t = 32$
to $N_t = 40$ gives longer plateaus in the effective mass plots so that 
hadron masses can be extracted more reliably.  We systemically study the
source size effects on the determination of the mass values.  Two non-local
$\Delta$ operators have also been added.  They are measured in both Coulomb
gauge and Landau gauge.

\section{SIMULATION}

Table~\ref{tab:parameters} lists the parameters of our calculation.
We use the 'R'-algorithm of Gottlieb, {\it et.\ al.} \cite{Gottlieb87} \,
for our evolution along with their notation.
We have collected 3100 microcanonical time units during a 5 month run 
on the 256-node Columbia parallel computer.  Hadron masses are measured
every 6 time units.  For a single gauge configuration, sources are set on 5
different time slices to perform an average over time slices (AOTS).
\begin{table}[hbt]
\vspace{-5mm}
\caption{Simulation parameters}
\label{tab:parameters}
\begin{tabular}{lc} \hline
	volume			&	$16^3 \times 40$	\\
	$\beta$			&	5.7		\\
	$m_{dynamical} a$	&	0.01		\\
	length			&	250--3100$\tau$	\\
	\\
	trajectory length	&	0.5$\tau$		\\
	step size		&	0.0078125$\tau$	\\
	CG stopping condition	&	$1.13 \times 10^{-5}$	\\
	minutes/trajectory	&	12.5		\\
	\\
	hadron source types	&	wall, 2Z 	\\
				&	wall, Z 	\\
	hadron source sizes	&	$16^3, 12^3, 8^3, 4^3$, point \\
	valence quark masses	&	0.004, 0.01, 0.015, 0.02	\\
				&	0.025, 0.05, 0.07	\\
	\hline
\end{tabular}
\vspace{-5mm}
\end{table}

For hadron mass calculations, our 2Z source is set
on a fixed $t$ slice and is non-zero when only all $(x, y, z)$ are even;
our Z source is non-zero for all $(x, y, z)$. 
Previously our 2Z and Z wall sources were non-zero across the entire
spatial volume.  In this run, however, we also set sources on smaller cubic
spatial volumes.  Besides the dynamical quark mass,
we also calculate quark propagators using different valence quark masses 
so that we can make hadron propagators out of different sets of quark masses.
Table~\ref{tab:parameters} also lists all of our hadron source sizes and 
valence quark
masses.  Out of all the possible combinations of source types, sizes and 
valence quark masses, we have a total of 19 combinations for mesons,
21 combinations the nucleon and 10 combinations for the $\Delta$.

\begin{figure}[htb]
\vspace{-10mm}
\epsfxsize=75mm
\epsffile{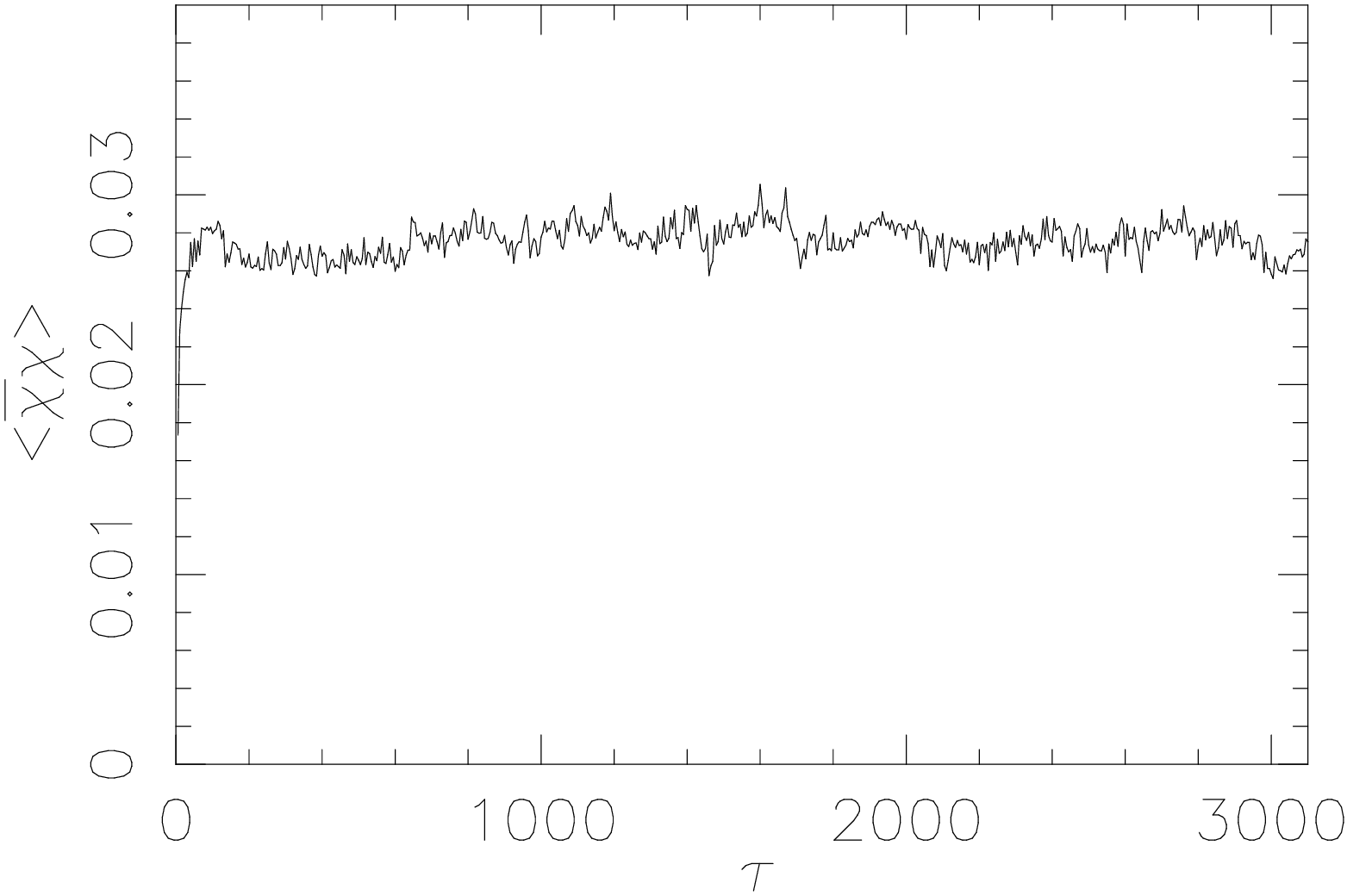}
\vspace{-12mm}
\caption{The evolution of the chiral condensate.}
\label{fig:evolution}
\vspace{-5mm}
\end{figure}

Figure~\ref{fig:evolution} shows the evolution of $\langle \overline{\chi}
\chi \rangle$ during our run.  Discarding the
first 250 time units for thermalization gives $\langle \overline{\chi}
\chi \rangle = 0.0276(2)$.  This is consistent with the value 0.0277(3) from
our earlier $16^3 \times 32$ run \cite{Hong91,Brown91}.

  \pagenumbering{arabic}
  \addtocounter{page}{1}

\section{RESULTS}

\begin{figure}[htb]
\vspace{-2mm}
\epsfxsize=75mm
\epsffile{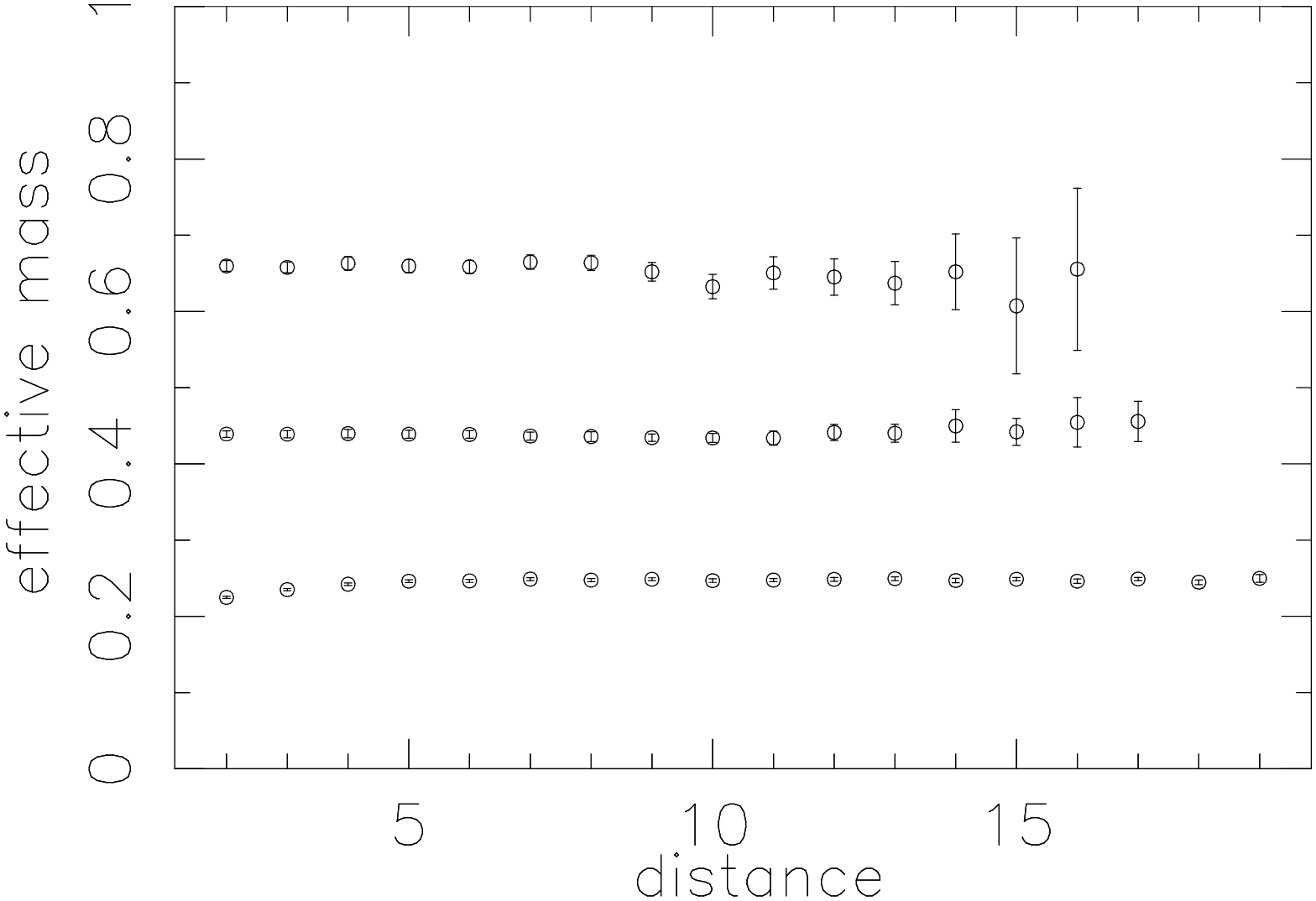}
\vspace{-12mm}
\caption{$N, \rho, \pi$ effective masses for $16^3$ source.}
\label{fig:effmass16}
\end{figure}

\begin{figure}[hbt]
\vspace{-7mm}
\epsfxsize=75mm
\epsffile{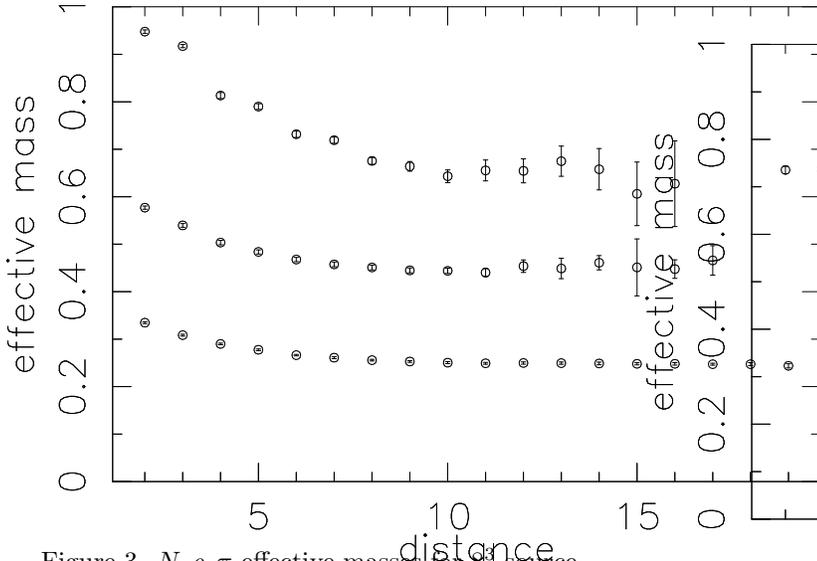}
\vspace{-12mm}
\caption{$N, \rho, \pi$ effective masses for $8^3$ source.}
\label{fig:effmass8}
\end{figure}

We see quite significant source size effects on the effective mass plots
for hadrons.  Figure~\ref{fig:effmass16} and ~\ref{fig:effmass8} show from top
to bottom the effective mass plots for $N, \rho, \pi$ on $16^3$ and $8^3$
sized sources.  For the $16^3$ source, the effective masses are quite flat as
the distance varies.  When the source size becomes smaller, the effective 
masses reach a plateau from above.  The smaller the size of the source,
the higher the value of the effective masses at short distances.
This behavior is seen with all our different sources: $4^3, 8^3, 12^3, 16^3$.
In fact, in our earlier $32^4, \beta=5.7, m a = 0.025$ calculation
with a $32^3$ source, all the effective masses rose from below and
the nucleon probably never reached a plateau.
For our current data, all the effective masses from 
different sized sources converge to the same plateau.
It is clear that $16^3$ is the optimal source size for this calculation because
it gives the longest plateau.


\begin{figure}[hbt]
\vspace{-2mm}
\epsfxsize=75mm
\epsffile{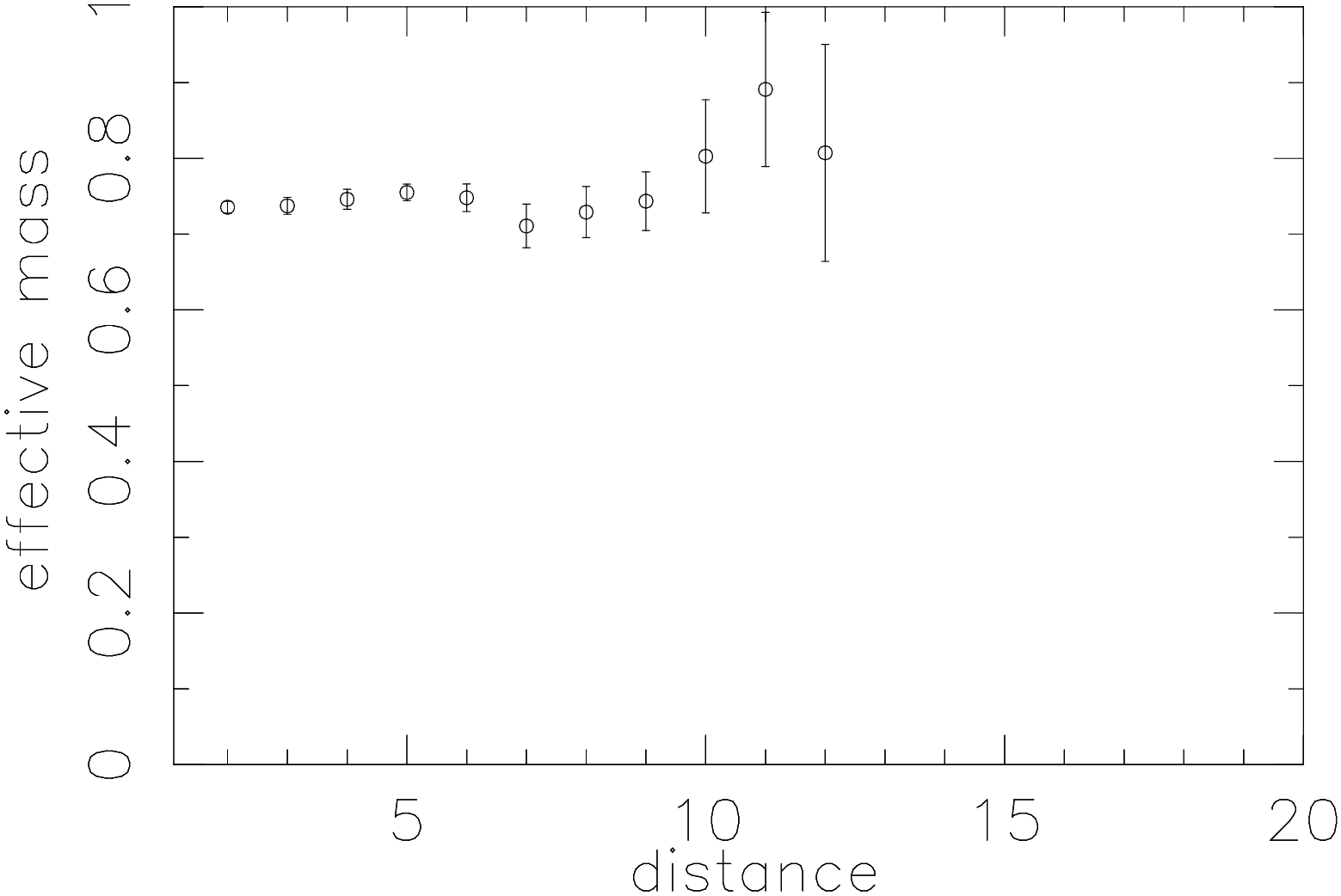}
\vspace{-12mm}
\caption{$\Delta_0$ effective masses for $16^3$ source.}
\label{fig:effmassd0}
\vspace{-5mm}
\end{figure}

Table~\ref{tab:masses} is a list of hadron mass values for our current 
$16^3 \times 40$ calculation.  Also listed for comparison
are our earlier $16^3 \times 32$ results.
All our fits are correlated fits with errors determined from the
jackknife method.  The pion mass is obtained from a single state fit with
the 2Z source.  All the other mesons and the nucleon come from
two state fits with the 2Z source.
Our $\Delta$'s come from two state fits with the Z source.  $\Delta_1$ and
$\Delta_0$ are two equivalent lattice operators corresponding to the two
$A_2$ operators in ref \cite{Golterman85}, respectively.
All our sinks are local except for the $\Delta$'s which are necessarily
non-local in the staggered fermion formalism.
Figure~\ref{fig:effmassd0} is an effective mass plot for
the Coulomb gauge $\Delta_0$ with a $16^3$ source. 

All our mass values are some what lighter than that found in the
earlier $16^3 \times 32$
run.  The nucleon is especially light, about 3 standard deviations lower than
the earlier result.  This is not because of different fitting ranges;
fitting our data with distances up to 16 instead of 20 gives
essentially the same
mass spectrum.  The two runs have about the same statistics.  The earlier one
has more averaging on a single configuration while the current one has more
independent configurations.  Except for the possibility that we are seeing 
a finite temperature effect (With antiperiodic boundary conditions in the time
direction, our larger $N_t$ lowers the temperature from 53 MeV to 44 MeV.),
one may argue that in both of the runs, the errors 
are underestimated and that our simulation time is not sufficiently large to
accurately compute the effects of autocorrelations.

The $\Delta$ masses measured in Landau gauge are quite different from those 
measured with both the source and the sink time slices in Coulomb gauge.
They are about 3 standard deviations lower.  
In fact, their masses are in closer agreement with the nucleon
than the Coulomb gauge $\Delta$'s.  This raises a question about
the validity of using Landau gauge, non-local operators in general,
since the gauge-dependent correlation between different time slices 
in Landau gauge may alter the signal one wants to obtain.

Setting the $\rho$ mass to 770 MeV to fix the scale, we find $m_{\pi}$ = 
436(6) MeV, $m_{\pi_2}$ = 502(9) MeV, $m_N$ = 1153(20) MeV and
$m_\Delta=1290(40)$ MeV (averaging over our Coulomb gauge $\Delta_0$ and
$\Delta_1$). Our lattice 
spacing is $a^{-1}$ = 1750(20) MeV and the lattice size is $L a$ = 1.80(2) fm.

\begin{table}[hbt]
\caption{Hadron mass results: $\beta = 5.7, ma = 0.01$.}
\label{tab:masses}
\begin{tabular}{lcc} \hline
  volume	&  $16^3 \times 32$	&	$16^3 \times 40$	\\
  length	&  1050--2700$\tau$	&	250--3100$\tau$		\\
  measured every&	5$\tau$		&	    6$\tau$		\\
  \# of measurements	&	330	&	    475			\\
  AOTS		&	32		&	      5			\\
  	\\
  $\langle \overline{\chi} \chi \rangle a^3$ & 0.0277(3) & 0.0276(2)	\\
  $m_{\pi} a$	&	0.252(3)	&	0.249(2)		\\
  $m_{\pi_2} a$	&	0.295(3)	&	0.288(4)		\\
  $m_{\rho} a$	&	0.454(4)	&	0.440(5)		\\
  $m_{\rho_2} a$ &	0.457(7)	&	0.440(6)		\\
  $m_{N} a$	&	0.692(6)	&	0.659(9)		\\
  $m_{N'} a$	&	0.795(10)	&	0.767(18)		\\
  $m_{\sigma} a$ &	0.441(14)	&	0.435(10)		\\
  $m_{A_1} a$	&	0.586(14)	&	0.576(15)		\\
  $m_{B} a$	&	0.610(9)	&	0.609(17)		\\
  $m_{N}/m_{\rho}$ &	1.527(11)	&	1.499(13)		\\
	\\
  $m_{\Delta_0} a$(C)	&		&	0.753(23)		\\
  $m_{\Delta_1} a$(C)	&		&	0.721(19)		\\
  $m_{\Delta_0} a$(L)	&		&	0.688(15)		\\
  $m_{\Delta_1} a$(L)	&		&	0.659(22)		\\
	\hline 	\\
\end{tabular}
The fits have $t_{min} = 9$ for the $\pi$ and $t_{min} = 6$ for all
other particles.  The $\chi^2/dof$ for all fits are $\leq 2$.
\end{table}

\end{document}